# Definition, Detection, and Recovery of Single-Page Failures, a Fourth Class of Database Failures

Goetz Graefe, Harumi Kuno

Hewlett-Packard Laboratories

Palo Alto, CA 94304

{Goetz.Graefe,Harumi.Kuno}@HP.com

## ABSTRACT

The three traditional failure classes are system, media, and transaction failures. Sometimes, however, modern storage exhibits failures that differ from all of those. In order to capture and describe such cases, single-page failures are introduced as a fourth failure class. This class encompasses all failures to read a data page correctly and with plausible contents despite all correction attempts in lower system levels.

Efficient recovery seems to require a new data structure called the page recovery index. Its transactional maintenance can be accomplished writing the same number of log records as today's efficient implementations of logging and recovery. Detection and recovery of a single-page failure can be sufficiently fast that the affected data access is merely delayed, without the need to abort the transaction.

## 1 INTRODUCTION

Modern hardware such as flash storage promises higher performance than traditional hardware such as rotating magnetic disks. However, it also introduces its own issues such as relatively high write costs and limited endurance. Techniques such as log-structured file systems and write-optimized B-trees might reduce the effects of high write costs and wear leveling might delay the onset of reliability problems. Nonetheless, when failures do occur, they must be identified and repaired.

Failure of individual pages on storage such as flash is not properly described by any of the traditional failure classes considered in transaction processing and database systems. Single-page failures are substantially different from transaction failures, from media failures, and from system failures. Among those three traditional failure classes, single-page failures are most similar to media failures. They differ from media failures, however, since only individual pages fail, not an entire device. Treating one or a few failed pages as a failure of the entire device seems very wasteful. In a system that relies on flash memory for all its storage, doing so would turn a single-page failure into a system-wide hardware failure. Similarly, in a traditional parallel server with single-disk nodes, a single-page failure would turn into a node failure. The problem in these cases is that today's techniques treat these failures as media failures. The introduction and definition of single-page failures as a fourth failure class, in addition to the three traditional failure classes, is the primary contribution of this paper.

For over 30 years, the three traditional failure classes have framed research and development in high availability and reliability of data storage. Extending the set of failure classes requires a strong conceptual motivation as well as a strong case for the practical value of the new failure class and its implementation techniques, including both detection and repair.

Detection and recovery techniques for single-page failures differ from those for the three traditional failure classes, not only in the algorithms but also in the degree to which recovery must disrupt transaction processing. For example, while media recovery usually takes many minutes or even hours and requires that all affected transactions be aborted, single-page recovery can be designed and implemented such that affected transactions merely wait a short time, perhaps less than a second, for repair and replacement of the lost page. Introduction of such a recovery technique, including a new data structure, its usage, and its efficient maintenance, is the second contribution of this paper.

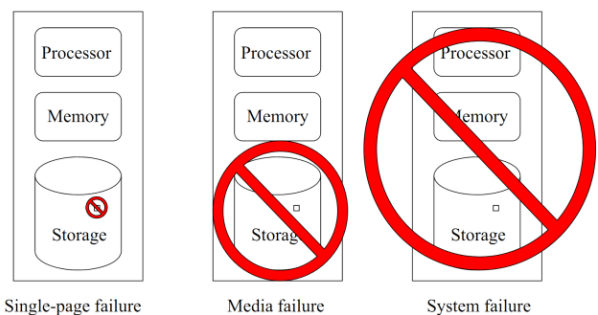

Figure 1. Failure scopes and possible escalation.

Figure 1 illustrates possible failure scopes. If single-page failures are not a supported class of failures, failure of a single page (left) must be handled as a media failure (center). In machines or nodes with only one storage device, a media failure is equal to a system failure (right). The present proposal aims to prevent such an escalation by introducing a new failure class including early detection and recovery of single-page failures.

While flash storage is known to wear out in individual locations while other locations are still perfectly viable, e.g., due to excessive local writing, the same problem already exists with traditional disk drives. In other words, single-page failures are not a new problem, merely one we may expect to get worse. As an example, a posting in an online forum [19] describes a real-world instance of the problem and its severity: "This one bit hard. Basically a disk started returning corrupted data for some sectors without actually failing the reads, so the controller didn't know anything was wrong and happily reported the raid5 array OK. It has therefore been doing parity updates based on misread info so by now pulling the disk won't help a bit since it'll just recreate the info that was misread. And since it was a silent failure, all backups are suspect as well since there's not really any way to know how long it's been going on. And even though we've paid for next-business-day on-site support, since the SAN was reporting the array OK, support claimed it isn't a hardware failure so now the computing cluster has been down for more than 2 weeks." Notice that errors "in some sectors" snowballed into a media failure (of the entire RAID-5 array and its backups) and then a system failure of an entire cluster (for weeks). As traditional disk storage is replaced by flash memory with known endurance limits, this kind of incident is liable to occur more often.

Bairavasundaram et al. [2, 3] gathered statistics from millions of disk drives and found "latent sector errors" in thousands of them, sometimes dozens or even thousands of errors in a single disk drive. Their definition of a latent sector error implies data loss, i.e., a disk unable to recover the page contents in spite of error correction codes and multiple attempts to read. Among "nearline" (SATA) disks, 9.5% suffered at least one latent sector error per year. A majority of those errors occurred in read operations and during "disk scrubbing," i.e., occasional re-reading of all disk pages to verify their contents by their checksums. Redundancy in RAID systems was able to recover some of the disk errors, but of course not all disks are protected by RAID or other redundant storage. And again, as flash storage replaces mature disk technology, the occurrence rates of such incidents probably will not get better.

The next section reviews existing techniques for detection and repair of single-page failures. The following three sections cover failure classes, failure detection, and failure recovery. Each of these sections covers first traditional techniques and then single-page failures. These sections are followed by thoughts on recovery performance for single-page failures. The paper ends with a summary and conclusions from this research.

## 2 RELATED EXISTING TECHNIQUES

In disk devices and operating systems, "bad block mapping" has long been a standard technique to cope with individual error-prone disk sectors. After writing a page, it is immediately "proof-read" and remapped if errors are detected. If proof-reading succeeds but a later read operation fails, however, neither disk devices nor operating systems can recover the page contents.

If multiple disks are used as a unit, redundancy can increase availability and reliability. For example, RAID-5 uses single redundancy to prevent data loss in any single disk failure and RAID-6 uses dual redundancy to prevent data loss in any dual disk failure. Nonetheless, most read operations employ only a single disk without checking the parity across the disk array. Thus, if a single disk produces erroneous data without indicating a failure, the application receives bad data and may eventually write bad data back. Note that the anecdote in the introduction included data loss in a disk array.

All commercial database systems include utilities to verify the integrity of a database, both logical integrity (integrity constraints and materialized views) and physical integrity (pages, indexes, and auxiliary data structures such as allocation bitmaps and compression dictionaries). The relationship between secondary indexes and primary indexes can be part of logical integrity (like a foreign key integrity constraint) or of physical integrity. These tools include Oracle's DBMS_repair and DB Verify, MySQL's myisamchk, IBM's db2dart (database analysis and reporting tool), as well as Sybase's and Microsoft's DBCC (database consistency check). Borisov et al. [1] introduce a software tool that invokes such tools "proactively and continuously" optimized for least cost and for timely notification of database administrators.

Most of these utilities run offline, i.e., prohibiting concurrent updates of affected tables or even the entire database, or are based on snapshot services provided by the file system or an equivalent software layer. In the latter case, updates are possible while the utility runs but immediately render the verification result out-of-date. Online checks are commonly used [17] but usually verify only the plausibility of a single page, e.g., a parity check, with no checks of relationships among pages. – In contrast, single-page failure detection as discussed in Section 4 encompasses continuous verification as side effects of query execution and index access. In order to make the case with a concrete data structure, this section focuses on B-trees [4, 6, 9], a ubiquitous storage structure in databases, information retrieval, and file systems. For some variants of B-trees, fast online checks can verify comprehensively all structural invariants, i.e., relationships among B-tree nodes and pages.

Commercial database systems also provide functionality to edit data pages, usually in a very low-level and error-prone format. The process requires careful construction of appropriate commands, which is laborious, risky, and inefficient. – In contrast, single-page recovery as discussed in Section 5 can be automatically invoked, automatically completed, and very fast.

We know of only one example of automatic page repair. It is implemented in SQL Server database mirroring, a feature formerly known as log shipping. If a page within a mirror is found to be inconsistent, it is automatically replaced by the corresponding page in the primary copy. If a page in the primary copy is inconsistent, it is frozen until the mirror has applied the entire stream of log records, whereupon the page is replaced by an up-to-date copy of the page from the mirror. Note that the recovery log is applied to the entire mirror database, not just the individual page that requires repair, and that the recovery process completely fails to exploit the per-page log chain already present in the SQL Server recovery log. – In contrast, the single-page recovery as discussed in Section 5 works within a single database, i.e., without keeping an entire mirror database current at all times.

The existence of these tools validates that occasional loss or corruption of a single disk sector or a single data page is a real-world problem. A sound, reliable, automatic, and efficient solution should be very valuable in products and services for database management, information retrieval, and file storage.

## 3 FAILURE CLASSES

As the essence of this paper is to extend the traditional three failure classes with a fourth one, this section reviews the traditional failure classes and then introduces and defines the new one. Subsequent sections discuss promising methods for failure detection and recovery, both for the traditional failure classes and for single-page failures.

## 3.1 Traditional Failure Classes

In some aspects, e.g., from a user's point of view, it is quite different whether a user requests that a transaction be terminated or the system decides to terminate a transaction, e.g., to resolve a deadlock. On the other hand, in order to conceptualize database failures and to cover all possible failures with appropriate recovery functionality, failures have traditionally been divided into three classes.

A transaction failure leaves other transactions running; only a single transaction fails and must roll back to ensure all-or-nothing semantics for each transaction.

A media failure focuses on a storage device; a typical example is a read-write head scratching a disk surface and rendering the entire disk useless. In such a case, the operating system and the database management system might continue running, but all transactions fail that have touched data on the failed media or attempt to do so later.

A system failure is most severe; the database management system and perhaps even the operating system require restart and recovery. All uncommitted transactions fail and recovery removes all their updates in the database; all committed transactions and their effects on the database can be preserved or repaired in order to ensure durability of transaction commits.

These failure classes have been defined and discussed frequently in the past; may a few references suffice here [7, 12, 13, 15, 18]. They are the foundation of today's failure detection, recovery, reliability, and availability.

## 3.2 Single-Page Failures

The class of single-page failures encompasses all failures to read a data page from storage correctly and with plausible contents despite all correction attempts in lower-level hardware or software, e.g., in the device firmware, the device controller, or the RAID controller. According to some, this type of error already happens "all the time"[1] with traditional disk drives, which typically forces replacement of the device and subsequent media recovery. New storage technologies such as flash memory are known to have limited endurance and therefore are likely to aggravate the problem.

The causes for single-page failure can be many, from temporary or permanent hardware malfunctions to delays or malfunctions in overloaded network-attached storage. By recognizing all such cases as a single class, a single detection mechanism and a single recovery mechanism can be designed, implemented, tested, and perhaps even verified in a more rigorous manner than traditional testing and quality assurance.

A failure of one or more individual pages is less severe than a failure of an entire device and thus quite different from a media failure. In systems with only one instance of a device type, failure of the entire device practically leaves the system without a device of that type, whereas failure of some individual pages leaves most of the device intact and operational. In particular, if a system relies entirely on flash memory as its one-and-only storage device, failure of this device means failure of the system and thus requires hardware replacement; whereas failure of merely some individual pages may be resolved by appropriate localized recovery. With appropriate recovery techniques, e.g., those outlined below, it is not even required that any transactions terminate. Instead, instant, focused, localized recovery can limit the "damage" to a small delay of the transactions attempting to access failed pages.

---

[1] James Hamilton, personal communication, Asilomar, CA, October 26, 2011.

Because single-page failure differs from all three traditional failure classes, it should become a fourth failure class that future transaction theory and practice should consider.

While the concept of single-page failures clearly pertains to contemporary storage devices based on traditional rotating magnetic disks or on flash memory, it might be argued that single-page failure is an oxymoron for future byte-addressable persistent storage such as non-volatile memory. The counter-argument, however, holds that organizing data structures, indexes, etc. in pages is crucial for fault containment including fault detection and efficient, localized repair. Sizes of these pages are undecided; they may, for example, be very large or vary in size within a single system or data store. Nonetheless, even if pages are not required for data movement in the memory hierarchy, the concept of pages remains a desirable building block for reliable software systems. If this counter-argument prevails in future data management using non-volatile memory as storage media, the concept of single-page failures seems to apply to all foreseeable computer systems.

## 4 FAILURE DETECTION

Necessary for correct automatic failure repair or recovery is reliable detection of failures. Failures might be detected within a single page (e.g., if some form of parity is used [17]), within a data structure such as a B-tree index, within a database table and its data structures (e.g., between primary and secondary indexes), within a single database (e.g., foreign key integrity constraints or consistency of materialized views and base tables), and beyond. Detection of errors or inconsistencies within each data page is simple and commonly included in standard processing. Errors among multiple pages are harder to detect, for example, inconsistencies among components of a complex data structure such as nodes in a B-tree index. The standard B-tree format does not support such functionality as a side effect of standard processing but some variants do. (More details are discussed below.) Inconsistencies between primary and secondary indexes are quite similar to inconsistencies between tables and materialized views. Many possible errors can be found as side effect of standard processing, but not all – an opportunity for further research.

## 4.1 Detection of Traditional Failures

Traditional failure detection algorithms run offline and scan the entire data store, e.g., an entire database. The algorithms in some systems still read many pages many times, whereas more scalable algorithms read each page only once. Moreover, the sequence in which pages are processed might not be material, which greatly improves efficiency as it reduces the number of seek operations for the storage device [14].

Offline algorithms require that the data store is in read-only mode. Thus, these are inherently disruptive to applications and user experience.

Some supposedly online algorithms require that the file system or equivalent low-level functionality in the data manager provide stable snapshots such that the algorithm for database verification and failure detection runs on stable, read-only data while applications and database runs updates by creating new page copies. Thus, when such verification algorithms complete, the final report is inherently and immediately out-of-date.

While necessary and helpful, such tests are not sufficient. These algorithms must be complemented by a verification of each page when it is first loaded from storage into the buffer pool.

## 4.2 Detection of Single-Page Failures

Many single-page failures may be discovered by in-page tests, e.g., parity and checksum calculations. In a database system, in-page tests may include analysis of all byte offsets and lengths in the page header and in the indirection vector pointing to individual records.

Earlier work has already covered failure detection, in particular verification of data structures in databases. The following is meant merely to convince the reader that comprehensive, incremental failure detection can be efficient and realistic in high-performance data management systems and their indexes.

As a concrete example, B-tree indexes are ubiquitous in databases, information retrieval, and file systems. For many implemented variants of B-trees, comprehensive online consistency checking is not possible or at least has not been invented yet. B-trees with fence keys instead of sibling pointers, however, can enable comprehensive verification as side effect of standard query processing, i.e., repeated B-tree searches with root-to-leaf passes. The only field in a B-tree node that cannot be verified, i.e., compared against a redundant value, is the PageLSN, i.e., the most recent log record that pertains to the data page, together with its location in the recovery log.

For comprehensive structural verification, each node requires a low and a high fence key, which are copies of the separator key posted in the node's parent when the node was split from its two neighbor nodes. Thus, branch nodes with N child pointers have N+1 key values instead of N key values as in $B^{link}$-trees [16] or $N-1$ key values as in the original B-tree design. Leaf nodes always contain at least two key values, of which one (for example the high fence key) must be an invalid record (also known as ghost record or pseudo-deleted record).

When following a pointer from a parent to a child, the key values next to the pointer in the parent must be equal to the fence keys in the child. This is true for all levels in a B-tree. For two nodes without common parent but with common grandparent or further ancestor node (also known as cousin nodes), all possible verification tests can easily be completed in two root-to-leaf passes to or through those cousin nodes, because the separator key in the common ancestor node is replicated in fence keys along the entire "seam" from this ancestor node to the leaf level. It is not required that these two root-to-leaf passes occur at the same time. The fence keys contain all information required for all structural verification of the B-tree.

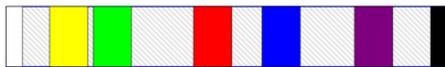
Figure 2. Symmetric fence keys in a page.

Figure 2 illustrates fence keys (shown white and black) in a B-tree node. All other key values fall between these two extreme colors. Due to suffix truncation (suffix compression) of separator keys in B-trees [5], the fence keys may be very small. Moreover, it might be convenient to include in one fence key the prefix truncated from all other key values in the page. The other fence key can alternate between valid record and ghost record in response to logical insertion and deletion of key values.

This set of invariants and incremental, instantaneous error detection can be combined with the advantages of $B^{link}$-trees [16]. Like $B^{link}$-trees, Foster B-trees [11] split nodes locally without immediate upward propagation; therefore, only two latches at a time suffice for all B-tree operations. Like write-optimized B-trees [8], Foster B-trees permit only a single incoming pointer per node at all times; therefore, they support efficient page migration and defragmentation as well as simple concurrency control during node removal. Due to symmetric fence keys [14], Foster B-trees permit continuous self-testing of all invariants; therefore, they enable very early detection of page corruptions.

Figure 3 shows a Foster B-tree with a root node, two more branch nodes, and three leaf nodes. Equal colors indicate equal key values. All nodes have two fence keys, i.e., copies of separator keys propagated to the appropriate parent during node splits. This is readily traced from root to right leaf. The left leaf node recently split, producing the center leaf node. For the time being, the left leaf node is the foster parent of the center leaf node, i.e., it acts as temporary parent node for its foster child. In a (temporary!) chain of foster relationships, each foster parent carries the high fence key of the entire chain. This permits comprehensive consistency checks along the entire chain starting with the parent node, e.g., the left branch node in Figure 3. Each time a root-to-leaf search follows a pointer, whether from a permanent parent to a standard child or from a foster parent to its foster child, the fence keys in the child must match two key values in the parent.

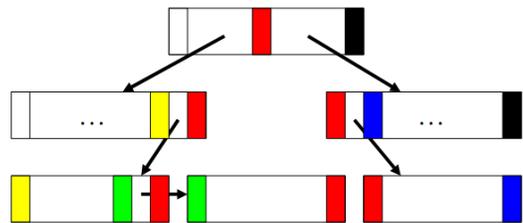
Figure 3. A Foster B-tree with a foster relationship.

The advantages of immediate detection are fairly obvious. For example, the nightmare example quoted in the introduction would have been impossible in a system testing all invariants, including both in-page parity and cross-page invariants. Instead, the system would have detected the failures upon their first occurrence and, ideally, been able to repair them using appropriate recovery techniques.

## 5 RECOVERY TECHNIQUES

The purpose of covering failure detection and recovery is to convince doubters that, with appropriate software design and implementation techniques, the definition of single-page failures as fourth failure class can have practical impact and advantages for database systems, applications, and users. We believe that better recovery techniques for single-page failures are possible and will evolve over time. The purpose here is merely to propose and outline an initial recovery technique that gives practical relevance to the new failure class.

All discussions of recovery techniques assume that the recovery log is on stable storage, i.e., once a log page has been written, it is not subsequently lost. While absolute certainty is unachievable, this assumption is found in practically all research and implementations of transactional data management systems. Typical implementation methods include writing multiple copies, perhaps on different types of devices in different locations. Data pages, on the other hand, including pages of the page recovery index (see below), may be lost in media failures and in single-page failures.

## 5.1 Traditional Recovery

Putting the proposed recovery technique for single-page failure into perspective requires a review of recovery techniques for traditional failures and of some specific optimizations that are in wide use even if they are not widely known. This section contains more detail on traditional techniques than prior sections, because these details have a strong relationship to the proposed single-page recovery technique. (This section is not supposed to provide an introduction or tutorial on database recovery techniques; an excellent source for that purpose is [18].)

### 5.1.1 Transaction Rollback

For efficient rollback of a single transaction, the log records of each transaction are linked together. Each log record points to the prior one, typically by means of an address in the recovery log or log sequence number (LSN). Let this be called the per-transaction log chain.

### 5.1.2 System Recovery

After a system failure, e.g., a crash of the operating system and all other software, database recovery employs three passes over the log: log analysis, "redo," and "undo." "Redo" is physical, i.e., applies to the same data pages, whereas "undo" is logical, i.e., applies to the same key values, e.g., in a B-tree index, and is often called "compensation" rather than "undo."

Log analysis and "redo" starts at the most recent checkpoint and go forward to the end of the recovery log; "undo" starts at the end of the recovery log and goes backward, going beyond the most recent checkpoint as much as required by transactions active during the checkpoint and incomplete at the time of the crash.

Log analysis and "redo" may be interleaved within a single pass over the recovery log. If they are separated, log analysis is very fast because it reads only the log but no data pages. If the analysis pass acquires locks, new transactions can be admitted into the system immediately after the log analysis is complete.

The "redo" pass must read all data pages with logged updates and ensure that these updates have been applied, which is decided by the log sequence number on the data page. These random reads in the database dominate the cost of the "redo" pass. Many of these random reads can be avoided if the recovery log indicates which pages have been written successfully to the database during normal processing and buffer cleaning.

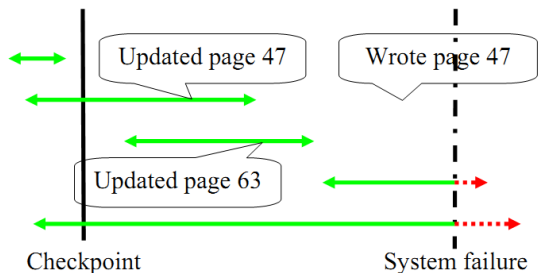

Figure 4. Optimized system recovery.

Figure 4, based on a widely copied diagram by Gray [7], illustrates the optimization. Assuming that a checkpoint writes all dirty data pages from the buffer pool and that the log records shown as callouts are the only log records for those data pages, recovery from the system failure needs to read and possibly update page 63 but it may avoid this effort for page 47. Note that a log record about a completed write operation is independent of any user transaction.

Log records for completed write operations, including the page log sequence number of each page written, are discussed but not recommended in the original ARIES paper [18]. They are used for fast system recovery in real systems, however, for example in Microsoft SQL Server. An optional optimization exploits any remaining free space in a log page by filling it with a list of recently written data pages and their log sequence numbers.

When a data page is reformatted (for whatever reason, but typically after allocation from the free space pool), it has the same effect as a successful write operation: "redo" for all prior log records is not required. If a backup copy of a data page is inserted into the recovery log (for whatever reason), the effect is very much like reformatting the page.

### 5.1.3 Media Recovery

Whereas system recovery scans the recovery log forward from the last checkpoint and ensures "redo" of all logged updates, media recovery scans forward from the last backup of the failed media and ensures updates for the failed media only. Due to the effort of restoring a backup copy, active transactions touching the failed media are aborted.

If the restore effort is also used to migrate to newer devices, the new media may employ different page identifiers, e.g., for parent-to-child pointers in B-tree indexes. Thus, restoring to alternative media requires remapping page identifiers. This task can be simplified and sped up if each page has exactly one incoming pointer at all times, e.g., in Foster B-trees [11].

### 5.1.4 Page Versioning

In addition to recovery techniques for the three traditional failure classes, the recovery log can also serve some concurrency control techniques. Specifically, snapshot isolation can be implemented by taking an up-to-date copy of a database page and rolling it back using "undo" information in the recovery log. This is the principal technique enabling snapshot isolation in Oracle's namesake database software.

An efficient implementation of single-page rollback requires that each log record points to the previous log record pertaining to the same data page. Let this be called the per-page log chain, which obviously is different from the per-transaction log chain. In addition to single-page rollback, a per-page log chain can also be used for other useful system features, e.g., recovery in shared-disk systems with local logs [20].

The per-page log chain serves also another purpose. During system recovery and "redo" of each log record, the log sequence number of the prior log record is also the expected previous log sequence number in the data page (PageLSN). This can be exploited to verify the correct sequence of "redo" actions during system recovery. Defensive programming seems a very good practice in general but in particular for code designed to repair a database after a crash. For this purpose, per-page log chains exist in many database systems, for example in Microsoft SQL Server.

### 5.1.5 System Transactions

For data structures that permit multiple representations for the same contents, e.g., B-trees, it is convenient to separate changes to database contents in user transactions and changes to their representation in system transaction. For example, a user transaction might require new free space for an insertion, but it invokes a system transaction to split the node and commit this split. System transactions are similar to top-level actions in ARIES [18].

System transactions are limited to contents-neutral structural updates of the database, e.g., splitting a B-tree node (without the insertion that triggered the split), compacting a page (to reclaim fragmented free space), or removing a ghost record (also known as pseudo-deleted record). Thus, system transactions do not require forcing the log buffer to stable storage. Their commit log records will be forced to stable storage prior to (or with) the commit log record of any dependent user transactions. Should a system failure prevent logging the commit log record of a system transaction, the system transaction is lost. Since the system transaction is, by definition, contents-neutral, a lost system transaction cannot imply any data loss.

|  | User transactions | System transactions |
|---|---|---|
| Invocation source | User requests | System-internal logic |
| Database effects | Logical database contents | Physical data structure |
| Data location | Database or buffer pool | In-memory page images |
| Parallelism | Multiple threads possible | Single thread |
| Invocation overhead | New thread | Same thread |
| Locks | Acquire and retain | Test for conflicting locks |
| Commit overhead | Force log to stable storage | No forcing |
| Logging | Full "redo" and "undo" | Omit "undo" in many cases |
| Recovery | Backward | Forward or backward |

Figure 5. User transactions and system transactions.

Figure 5, copied from [10], summarizes the differences between user transactions and system transactions. The principal value of system transactions is their low overhead; their principal limitation is that they must not modify logical database contents.

### 5.1.6 Summary: Traditional Recovery

In summary, traditional recovery techniques cover the three traditional failure classes. Common optimizations include logging successful writes of data pages and per-page log chains.

## 5.2 Single-Page Recovery

In order to lend practical value to a new failure class in addition to the traditional three failure classes, recovery of such failures is required. The present section proposes one such method. Some of its techniques resemble optimizations of prior recovery techniques, in particular per-page log chains and logging successful writes. Nonetheless, the proposed recovery technique requires a new data structure in each database, which we call the "page recovery index." In addition, it relies on earlier versions or backup copies of individual database pages. Both are discussed below.

While single-page recovery is described below for only one page, it is perfectly possible that multiple pages fail and that they be recovered at the same time. The failed pages might be contiguous or not, and there might be very few or very many failed pages. In the case of multiple single-page failures, their recovery might be coordinated, e.g., with respect to access to the recovery log; we omit those variants here even if needed in a full implementation.

Suffice it to say that if all pages on a storage device require recovery at the same time, and if their recovery is coordinated, then access patterns and performance of the recovery process resemble those of traditional media recovery.

The techniques described below assume that the page recovery index is perfectly reliable. If a read operation fails in the page recovery index or produces page contents that fail consistency tests, then it is always possible to treat the failure as a media failure. Without doubt, a complete implementation would apply single-page recovery to the page recovery index as well.

In order to preempt any confusion: the page recovery index might be a B-tree and online comprehensive failure detection was discussed using B-trees as an example data structure, yet the recovery techniques discussed below apply to any storage structure.

### 5.2.1 Sources of Backup Pages

Efficient recovery from a single-page failure requires an earlier copy of the failed page. The most obvious source for an old copy is a database backup, i.e., the same type of archive copy as required after a media failure. As during recovery from a media failure, the recovery log is also required covering the time from the backup to the present time. The difference to media recovery is that single-page recovery requires only an individual data page. Thus, a single, a sequentially compressed backup image of an entire database is less than ideal.

Alternatively, normal transaction processing might occasionally take copies of data pages. This might occur explicitly with a view to possible single-page recovery. For example, a conservative policy might take such a copy after every 100 updates of a data page. Note that taking copies of frequently updated data pages takes less space than a traditional differential backup, because these backups need space only for pages with many updates rather than for pages with any updates.

Page copies might also remain after a page migration. In a log-structured file system or a write-optimized B-tree (which allocate a new page location for each write), this means merely deferring space reclamation. Moreover, wear leveling, defragmentation, or index reorganization move database pages. The old, pre-move image might be retained and serve as single-page backup.

In some cases, the recovery log itself might serve as the source of a backup image. For example, when a page is formatted after allocation from the free page pool, the log record containing formatting information for the initial page image may substitute for an explicit backup copy. Similarly, taking an occasional full-page image (presumably compressed) of a database page in the recovery log supplants the need for a backup copy.

For each database page, the page recovery index tracks the most recent backup location.

### 5.2.2 Page Recovery Index

The page recovery index, one for each database or table space with an entry, tracks two pieces of information required for recovery from single-page failures. For each data page, it retains information about the most recent backup copy and, if the data page has been updated since the most recent backup, the log sequence number (location) of most recent log record pertaining to the page. While a data page is present in the buffer pool, its entry in the page recovery index may fall behind, but while a data page is not present in the buffer pool, its entry in the page recovery index maps the page identifier to the most recent log record pertaining to the data page (its PageLSN).

Thus, after a page has been updated and then evicted from the buffer pool, the page recovery index requires an update reflecting the new PageLSN in the data page. If the page lingers in the buffer pool after the write operation, the update of the page recovery index must occur sometime between the write operation and the eviction from the buffer pool. Similarly, the information about backup copies is updated when a backup copy is taken for a data page, when the page moves and the old version is kept as backup copy, or when a page is formatted (after allocation from free space) and all formatting information is logged.

Figure 6 illustrates the situation after an update of a data page in the buffer pool. The chain of log records is anchored by the PageLSN in the data page and embedded in the log records as per-page log chain. The page recovery index is not maintained as part of the data update; therefore, its information about the most recent log record is not reliable (shown as dashed line). Maintenance of the page recovery index occurs only after the dirty data page has been written back to the database. The situation after writing the data page, maintenance of the page recovery index, and eviction from the buffer pool is shown in Figure 9 and discussed shortly.

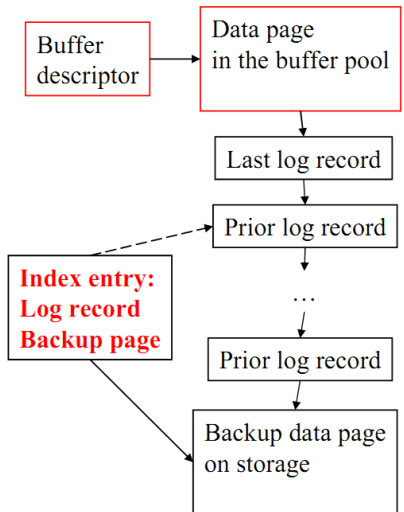

Figure 6. Data structures per page – data page buffered.

The PageLSN in the page recovery index serves recovery after a single-page failure. It indicates the start of the chain of log records required for bringing a backup page up-to-date. If the page backup is a page retained in a page movement such as defragmentation, the recovery log probably even contains a reference to the movement and the backup page. In other words, there are cases in which the identifier of the backup page is available via the PageLSN.

Even if the page identifier of the backup page might seem superfluous, it serves an important purpose. When a new backup page is taken during normal forward processing, the old backup page may be freed and the page recovery index gives fast access to its identifier. For an instant, the old and the new backup pages exist concurrently. It is not a good idea to overwrite an existing backup page, because the backup and recovery functionality are lost if this write operation fails.

Figure 7 summarizes the information in the page recovery index. The information on the backup page can be one of those three alternatives. The log sequence number information in the page recovery index is not maintained while the data page is in the buffer pool. When a page is read into the buffer pool, comparing the PageLSN in the data page with the information in the page recovery index is an additional consistency check that could prevent the nightmare recounted in the introduction. Note that the PageLSN was singled out in Section 4.2 as the only field in a B-tree node that cannot be verified as side effect update and query processing – the page recovery index resolves this insufficiency.

| Field | Contents | Comments |
|---|---|---|
| Backup page | Page identifier or log sequence number of last page formatting or of in-log copy | Used when freeing the old backup page when taking a new page backup. |
| Log sequence number | Most recent page update | Valid only if the page is not resident in the buffer pool and has been updated since the last backup. |

Figure 7. Fields in the page recovery index.

Figure 8 illustrates the logic for reading a page after a buffer fault. If a newly read data page fails the appropriate consistency check, a traditional system offers no choice but declare a media failure. If single-page failures are supported, the system keeps running using a single-page recovery procedure.

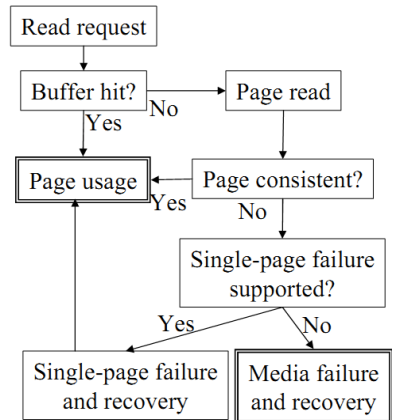

Figure 8. Page retrieval logic.

For the page recovery index, an ordered index (as opposed to a hash index) permits the best compression. For example, a single entry should cover a large range of pages if they all have the same mapping, e.g., a backup of the entire database (for the purpose of single-page recovery, the backup should be on direct-access media, e.g., disk rather than tape). If only one page within such a range is given a new backup page, the range must be split as appropriate. In the opposite extreme, there might be as many records in the page recovery index as there are pages in the database. In the worst case, the size of the page recovery index may reach about 16 bytes per database page or about 1‰ of the database size. Thus, it seems reasonable to keep the page recovery index in memory at all times.

Since the page recovery index is stored in database pages, its pages are covered by its contents. In order to prevent a data page containing information required for its own recovery, the database and the page recovery index might each be divided into two pieces such that the one piece of the page recovery index is stored in one piece of the database yet covers all data pages in the other piece of the database. Other schemes may be more appropriate for a specific implementation context.

### 5.2.3 Recovery Procedure

While a page is in the buffer pool, that copy is kept up-to-date. When a page is read into the buffer pool, a single-page failure might be detected. In that case, single-page recovery first retrieves information from the page recovery index and restores the backup copy into the buffer pool. The backup copy might be a log record describing the initial contents of the page immediately after it was newly allocated from the free space pool.

Using the log sequence number obtained from the page recovery index, single-page recovery follows the per-page log chain (see Section 5.1.4) back to the time the backup was taken, pushes pointers to those log records into a last-in-first-out stack, and then pops records off the stack and applies their "redo" actions.

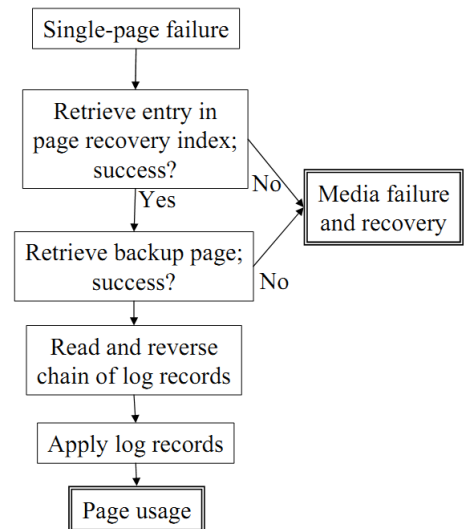

Figure 10. Single-page recovery logic.

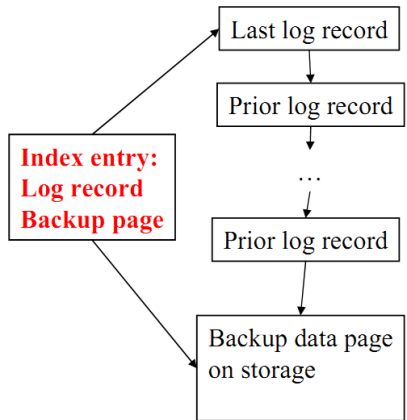

Figure 9. Data structures per page – ready for recovery.

Figure 9 illustrates the data structures required for single-page recovery, including the contents and the role of the page recovery index. For any data page, the entry in the page recovery index points to the most recent backup page and the most recent log record. The former might be an individual copy of the data page, a page within a database backup, or a log record about allocation and formatting of a new page. The latter is required only if the database page has been updated since the last backup. During recovery, the chain of log records is applied to the backup in order to obtain the latest, up-to-date page contents.

Figure 10 illustrates the steps in a single-page recovery. If anything fails, e.g., retrieval of an appropriate entry in the page recovery index, the system can resort to a media failure and appropriate recovery, of course with the implied disruption of transaction processing. Much more often, single-page recovery can be expected to succeed using a page back and appropriate log records linked together in the per-page log chain.

Once the page contents has been recovered and brought up-to-date in the buffer pool, the page can be moved to a new location. The old, failed location can be deallocated to the free space pool or registered in an appropriate data structure to prevent future use (bad block list). Obviously, differently from the process in other page migrations, the failed page must not be recorded as a backup page in the page recovery index.

### 5.2.4 Maintenance of the Page Recovery Index

Updates of the page recovery index are required when formatting a page after allocation, when taking a backup copy, or when evicting an updated page from the buffer pool. Better than during eviction, the update of the page recovery index immediately follows the write operation that turned a dirty data page in the buffer pool into a clean one. (Writing and eviction often coincide, but not always.)

As with updates in standard indexes, it is not required to force updates in the page recovery index to the database. It is, however, desirable to log those updates such that they can be applied later if necessary. If eviction of newly cleaned pages in the buffer pool is not urgent (due to contention), the buffer pool might retain a data page including its PageLSN until the log record for the page recovery index is saved on stable storage.

Logging updates in the page recovery index is an effort in addition to the database updates requested by user transactions. It turns out, however, that these log records written after a page has been written from the buffer pool to the database are very similar to those discussed earlier in Section 5.1.2 as a logging optimization for faster restart recovery. After each completed page write follows a single log record. The page recovery index subsumes the value of logging completed writes.

This leaves the question whether those log records must be forced to stable storage as part of a transacted update of the page recovery index. Doing so would add a forced log write to each database write; clearly a very high cost. While each update of the page recovery index could and should be a transaction, it could be treated as a system transaction, which does not require forcing the log upon commit [10].

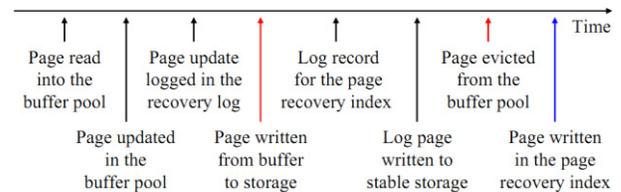

Figure 11. Update sequence for the page recovery index.

Figure 11 illustrates the individual update states in a database and its page recovery index. After a dirty page in the buffer pool is written back to the database, a log record describing the appropriate update in the page recovery index is written before the data page is truly evicted and replaced in the buffer pool. After this log record has been saved in the log, there is no urgency to write the data page of the page recovery index.

### 5.2.5 System and Media Recovery

During system recovery, specifically during log analysis, log records describing updates in the page recovery index also imply successful writes of data pages. Thus, these log records enable the same speed-up of the "redo" phase as logging successful writes as described in Section 5.1.2.

More interesting are the cases in which a data page was written successfully but a system or media failure prevented an update of the page recovery index or writing the appropriate log record to the recovery log. In such cases of lost updates in the page recovery index, the lost updates must be repaired, which will require precisely the same random reads required in "redo" recovery even if successful writes are logged.

If an update to the page recovery index is lost in a system failure, the case can easily be detected and repaired during system recovery. Just as in today's optimized recovery from a system failure described in Section 5.1.2, the log analysis pass will find a log record for an update in the data page but not log record for an update of the page recovery index. Thus, recovery must assume that the data page was not written prior to the system failure. The "redo" pass will read the data page, inspect its PageLSN, and determine whether the log record is already reflected in the data page. If not, recovery must re-apply the update from the log record. The page recovery index will be updated when the up-to-date data page is written back to the database during recovery or a subsequent checkpoint.

| Recovery phase | Log record type | Recovery action |
|---|---|---|
| Log analysis | Update a data page | Add the data page and this LSN to the recovery requirements |
| | Update an entry in the page recovery index | Remove the data page from the recovery requirements; add the page in the page recovery index |
| "Redo" | Update a data page (no matching update in the page recovery index) | Read the data page and check its PageLSN; if lower than the present LSN, update the data pages; otherwise, create a log record for the page recovery index |

Figure 12. Recovery actions.

Otherwise, the data page already reflects the log record, i.e., the data page had been written back to the database prior to the system failure. The system failure must have preempted the update to the page recovery index and the corresponding log record. In that case, the page recovery index must be updated right away, because there is no guarantee that a subsequent update will force an update of the page recovery index. To be more precise, the recovery process should generate an appropriate log record for an update of the page recovery index, but just as during normal transaction processing, there is no need to apply this update to the page recovery index immediately.

Figure 12 summarizes the actions required during recovery from a system failure. The log analysis phase simply determines which data pages require recovery, including data pages in the page recovery index. The "redo" phase applies the recovery actions and brings the page recovery index up-to-date if necessary.

### 5.2.6 Checkpoints

Database checkpoints force dirty data pages from the buffer pool to the database and log buffers to stable storage. This naturally includes data pages of the page recovery index as well as appropriate log records. However, as data pages are written out, the page recovery index is updated and appropriate new log records are created. Thus, there seem to be a never-ending tail writing data pages, updating the page recovery index, writing those newly updated pages of the page recovery index, etc.

Instead, it seems sufficient to write only those data pages and log records of the page recovery index that were dirty when the checkpoint operation started. All subsequent updates of the page recovery index will, in the worst case, create a small number of random read operations during a system recovery, should system recovery be required shortly after the checkpoint.

### 5.2.7 Summary: Single-Page Recovery

The described proposal for recovery from single-page failures, including its overall efficiency and online behavior, validates that this new failure class can be of practical use. If a single-page failure occurs, it can be detected and repaired so efficiently that it is not required to terminate the affected transaction. Instead, a short delay equal to a few I/O operations suffices to recover the desired contents of the data page.

## 6 PERFORMANCE EXPECTATIONS

While the design for failure detection and recovery is designed merely to lend credibility to the primary contribution of our research, namely the definition of single-page failures in database recovery, the following considerations may illustrate the performance expectations for single-page recovery and how much it differs from media recovery.

The three traditional failure classes and their recovery techniques differ widely in their expected recovery times: transaction rollback typically takes less than a second, system recovery should take about a minute depending on checkpoint frequency, and media recovery can take hours even after installation of a replacement device. For example, restoring a backup with 100 GB of data at 100 MB/s requires 1,000 s or about 17 minutes. Restoring a modern disk device of 2 TB at 200 MB/s requires 10,000 s or about 3 hours. Replaying recent log records, i.e., all log records since the backup operation, adds further time to media recovery.

The duration of single-page recovery using the proposed algorithms and data structures is probably closest to that of transaction rollback. It may take dozens of I/Os in order to read the required log records in the recovery log plus one I/O for the backup page. Thus, pure I/O time should perhaps be 1 s. Reversing the

654

sequence of log records with a last-in-first-out stack is practically free and applying dozens of log records in memory should also be very fast. Thus, the total time for recovery from a single-page failure should be a second or less. This delay can be absorbed within a transaction, i.e., without transaction abort even for an interactive user transaction.

Fast single-page recovery can be ensured with a page backup after a number of updates or after a period since the last page backup. To do so, the age of the last page backup can be kept in the page recovery index. The number of updates can be counted within the page, incremented whenever the PageLSN changes. The number of log records that must be retrieved and applied to the backup page equals the number of updates since the last page backup.

## 7 SUMMARY AND CONCLUSIONS

In summary, single-page failures are a fourth failure class that, like the traditional three failure classes, has practical importance, can be detected when it occurs, and permits efficient recovery. It is different from media failure as it pertains to individual pages, not to entire devices. Moreover, recovery of a single page does not require termination of any transaction, because the recovery procedure can retrieve and apply log records very selectively and efficiently.

With various modern hardware technologies, including flash storage and non-volatile memory, single-page failures are a concern. One means to reduce this concern is wear leveling, e.g., in the flash translation layer. However, wear leveling and related techniques can only reduce their probability or delay their onset but they cannot reliably prevent single-page failures.

The proposed recovery technique requires a new data structure in each database, the page recovery index. The logging effort for the page recovery index can be negligible as it is equal to the effort for logging completed writes, which some real systems already do in order to speed up system recovery should it become required. The effort for system and media recovery with and of the page recovery index is similar to that of recovery in traditional systems that log completed writes.

In conclusion, the class of single-page failures provides a powerful abstraction for various possible faults of traditional and modern hardware. It enables better reliability and availability of storage technologies that occasionally incur localized data loss. Without doubt, future designs and implementations will provide better recovery procedures than our design.

## ACKNOWLEDGEMENTS

Gary Smith suggested comparing the PageLSN of a page newly read into the buffer pool with the information in the page recovery index. Bernhard Seeger's insightful comments and suggestions improved the presentation of the material. The anonymous reviewers also gave numerous helpful suggestions.